\documentclass[journal]{IEEEtran}

\usepackage{algorithmic}
\usepackage{cite}
\usepackage{graphicx}

\title{Improved construction of irregular progressive edge-growth Tanner graphs}

\author{
  \IEEEauthorblockN{Jesus Martinez-Mateo, David Elkouss, and Vicente Martin}
\thanks{
The authors are with the Facultad de Inform\'atica, Universidad Polit\'ecnica de Madrid, Madrid, Spain (e-mail: \{jmartinez, delkouss, vicente\}@fi.upm.es).}
}

\begin{document}
\maketitle

\begin{abstract}
The progressive edge-growth algorithm is a well-known procedure to construct regular and irregular low-density parity-check codes. In this paper, we propose a modification of the original algorithm that improves the performance of these codes in the waterfall region when constructing codes complying with both, check and symbol node degree distributions. The proposed algorithm is thus interesting if a family of irregular codes with a complex check node degree distribution is used.
\end{abstract}

\begin{IEEEkeywords}

Irregular low-density parity-check codes, progressive edge-growth algorithm, waterfall region.

\end{IEEEkeywords}

\IEEEpeerreviewmaketitle

\section{Introduction}

Low-density parity-check (LDPC) codes were introduced by Gallager in the early 1960s~\cite{Gallager62}. These codes are capacity achieving for many communication channels, and though forgotten for years, are nowadays ubiquitous as they have been found useful in many communications problems.

A family of LDPC codes is usually defined by two generating polynomials, $\lambda(x)$ and $\rho(x)$. The coefficients of these polynomials define the distribution of incident edges to symbol and check nodes respectively. Richardson \emph{et al.} showed that the asymptotic behaviour of a family of codes defined by both polynomials can be analysed by using the density evolution algorithm~\cite{Richardson01}. Families of LDPC codes performing close to the channel capacity can be then designed by optimising both generating polynomials~\cite{Shokrollahi00, Elkouss09}.

An LDPC code is a linear code identified by an sparse parity-check matrix or its equivalent bipartite graph, also called Tanner graph. It is known that some iterative algorithms, such as belief propagation based algorithms, provide optimum decoding over cycle-free Tanner graphs~\cite{Tanner81}. However, any finite-length graph has necessarily cycles, and it has been shown that a large girth (length of the shortest cycle) improves the performance of LDPC codes using iterative decoding as it enforces a reasonable minimum distance~\cite{Hu05}. Therefore, taking into account that a finite-length graph has cycles, it is important to make its girth as large as possible.

The progressive-edge-growth (PEG) algorithm is an efficient method for constructing Tanner graphs with large girth~\cite{Hu05}, in most cases with better performance than randomly constructed codes. PEG algorithm's interest lies in its simplicity, and its flexibility when constructing codes from a complex symbol node degree distribution. However, note that a large girth does not automatically imply a a large minimum distance. The performance of these codes can be improved, for instance, in the error floor region. Significant research in PEG-based algorithms has been done to achieve this improved performance.

The rest of this paper has been organised as follows. In Section II, a modified PEG algorithm is proposed for the construction of LDPC codes following both $\lambda(x)$ and $\rho(x)$ degree distributions for symbol and check nodes, respectively. In Section III, simulation results are shown. These results are compared with the original PEG algorithm and similar alternatives. In Section IV, the modified PEG algorithm and its applications in the construction of new codes are discussed.

\section{Modified PEG Algorithm}
\label{sec:algorithm}

A PEG-based algorithm consists of two basic procedures: a local graph expansion and a check node selection procedure. Both procedures are executed sequentially in order to construct a Tanner graph connecting symbol and check nodes in an edge-by-edge manner. In the first procedure it is performed the expansion of the local graph from a symbol node, this expansion is used to detect and avoid short cycles when adding a new edge. The result is that check nodes that will produce a cycle are pruned, or if it is not possible to avoid a cycle, there only remains a set of candidate check nodes producing the largest cycle. The selection procedure is used to reduce this list of candidate nodes according to the current graph setting. In typical PEG algorithms, this procedure attempts to balance the degree of any check node selecting those candidates with the lowest check node degree.

Recent research has been focused on the case when there are several candidate nodes after both procedures. At this point, it is possible to improve the performance of a PEG-based algorithm, for instance avoiding small stopping sets for the binary erasure channel (BEC)~\cite{Kim07, Lin08, Xiaopeng09} or trapping sets in the binary symmetric channel (BSC). New definitions have also been introduced, such as the extrinsic message degree (EMD) or the approximate cycle EMD (ACE)~\cite{Tian03, Hua04}, which are two common measures used to calculate the connectivity of symbol nodes.

In the original PEG algorithm~\cite{Hu05}, a code is constructed according to a symbol node degree sequence. This sequence is previously calculated with the number of symbol nodes $n$ and the edge degree distribution established by $\lambda(x)$. Note that the original proposal does not take into account the second polynomial, $\rho(x)$, for the check node degree distribution. The algorithm proposed in this paper follows both degree distributions during the code construction procedures, changing the edge selection criterion, thus obtaining a better approach to codes with an irregular degree distribution.

\begin{figure}
\includegraphics[width=\linewidth]{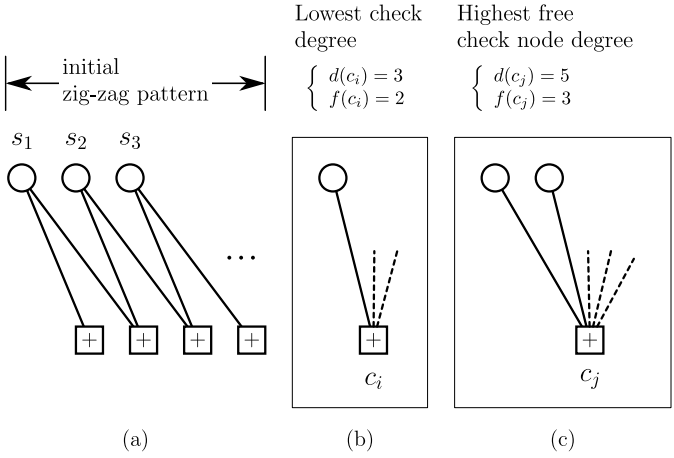}
\caption{Tanner graphs illustrating some characteristics of the algorithm. In (a) the zig-zag pattern used for 2-degree symbol nodes is shown. (b) and (c) show check nodes with different degrees, $d(c_i)$ and $d(c_j)$, and different number of incident edges (partially assigned), $d^k(c_i) = 1$ and $d^k(c_j) = 2$. First check node, $c_i$, is selected if a lowest check node degree criterion is used, while the second node, $c_j$, is selected if the criterion used is the FCD (see Section~\ref{sec:criterion-fcd}).}
\label{fig:criterion}
\end{figure}

\subsection{Free Check-node Degree (FCD) Criterion}
\label{sec:criterion-fcd}

The edge selection procedure used in this proposal differs from the selection procedure proposed in the original PEG algorithm. The graph is analysed to avoid local short cycles, however check nodes are not chosen according to its number of assigned edges, $d^k(c_i)$, i.e. its current (or partial) degree.

Instead of this, the check node with the highest difference between its partial and final-defined degree is chosen, $f(c_i) = d(c_i) - d^k(c_i)$, i.e. the difference between the number of currently assigned edges and the total number of edges to be assigned. The lowest check node degree procedure is replaced by a highest free check node degree (FCD) procedure (see Fig.~\ref{fig:criterion}). The FCD concept, comes from the concept of ``sockets'' previously described in~\cite{Richardson01, Richter05}.

We introduce the concept of \textit{compliance} of a constructed code as the distance between the distribution of nodes (symbol or check nodes) in the code and the pre-established node degree distribution. Let $\rho_j$ be the pre-established probability distribution for a degree $j$ check node, and $\rho_j^*$ the actual probability of a degree $j$ check node in the constructed graph, we calculate the $\rho$-compliance of a code as:

\begin{equation}
\label{eq:compliance}
\eta = \sum_{j=2}^{d_{c_j}^{\max}} \left| \rho_j -  \rho^*_j \right|
\end{equation}

\noindent where $\rho^*(x) = \sum_{j=2}^{d_{c_j}^{\max}} \rho_j^* x^{j-1}$, and $d_{c_j}^{\max} = \max \{ d(c_j) \}$.

\subsection{Modified Progressive Edge-Growth (PEG) Algorithm}
\label{sec:algorithmic-description}

A modified PEG algorithm is described below.

\begin{algorithmic}

\REQUIRE $d(s_i) \leq d(s_j)$ $\forall i < j $ and $f(c_i) = d(c_i)$ $\forall i$

\FOR{$j = 1$ to $n$}

\FOR{$k = 1$ to $d(s_j)$}

\IF{$k = 1$}

\IF{$d(s_j) = 2$}

\STATE $E_{s_j}^1 \leftarrow (c_i, s_j)$, where $E_{s_j}^1$ is the first edge incident to $s_j$ and $c_i$ is a check node such that it has the \emph{lowest check-node degree} under the current graph setting $E_{s_1} \cup E_{s_2} \cup \cdots \cup E_{s_{j - 1}}$.

\ELSE

\STATE $E_{s_j}^1 \leftarrow (c_i, s_j)$, where $E_{s_j}^1$ is the first edge incident to $s_j$ and $c_i$ is a check node such that it has the \emph{highest free check-node degree}.

\ENDIF

\ELSE

\STATE Expand a subgraph from symbol node $s_j$ up to depth $l$ under the current graph setting, such that $\mathcal N_{s_j}^l = \mathcal N_{s_j}^{l + 1}$, or $\overline{\mathcal N}_{s_j}^l \neq \emptyset$ but $\overline{\mathcal N}_{s_j}^{l + 1} = \emptyset$.

\STATE $E_{s_j}^k \leftarrow (c_i, s_j)$, where $E_{s_j}^k$ is the $k$th edge incident to $s_j$ and $c_i$ is a check node picked from the set $\overline{\mathcal N}_{s_j}^l$ having the \emph{highest free check-node degree}.

\ENDIF

\STATE $f(c_i) = f(c_i) - 1$

\ENDFOR

\ENDFOR

\end{algorithmic}

The same notation as in~\cite{Hu05} is used, $d(s_j)$ is the $s_j$ symbol node degree (The number of incident edges. It corresponds to the cardinality of the ensemble $E_{s_j}$ after the code construction), $d(c_i)$ is the $c_i$ check node degree, $f(c_i)$ is the number of edges that can be added to the check node $c_j$ under the current graph setting, such that $d^k(c_i) = f(c_i) - d(c_i)$, $E_{s_j}$ is the ensemble of edges connected to the symbol node $s_j$, $E_{s_j}^k$ the edge added in the step $k$ of the progressive construction, and $\mathcal N_{s_j}^l$ is the ensemble of nodes reached after the graph expansion from the symbol node $s_j$ up to depth $l$.

A zig-zag construction for 2-degree symbol nodes (see Fig.~\ref{fig:criterion}) is forced by using a special criterion when adding the first edge to a symbol node. In this particular selection, a list of eligible check nodes is limited to those check nodes already connected under the current graph setting, $E_{s'} = E_{s_1} \cup E_{s_2} \cup \cdots \cup E_{s_{j-1}}$, i.e. to the list of check nodes that have been chosen at least once from the first to $j$-th step. This construction is used to avoid cycles with 2-degree symbol nodes, thus obtaining a good performance in the error floor region as the results show (see Section~\ref{sec:results} below).

\subsubsection*{Relaxed edge selection}

The proposed PEG algorithm can be modified to work with a relaxed edge selection. In this case, if there are not check nodes with free edges in the final ens\-emble of candidate check nodes, $\overline{\mathcal N}_{s_j}^l$, check nodes with free edges are searched in the previous candidate ensemble, $\overline{\mathcal N}_{s_j}^{l-1}$. This procedure improves the $\rho$-compliance, $\eta$, with the target check node degree distribution, $\rho(x)$, at the expense of the current local cycle length (see Table~\ref{tab:rho-compliance}).

\section{Simulation Results}
\label{sec:results}

Simulations results have been done for different LDPC code construction methods. All constructed codes have a codeword length $n = 10^5$ and rate one half, $R = 0.5$. Performance has been measured under iterative decoding by using belief propagation. The maximum number of iterations for the decoder was set to 2000. Five different PEG-based construction methods are compared: (1) the original PEG algorithm as proposed in~\cite{Hu05}; (2) the modified PEG algorithm proposed by Richter in~\cite{Richter05}; (3) a modification of (2) by inforcing the selection of a check node in the current graph when the first edge is added to a symbol node; (4) the modified PEG algorithm proposed here; and (5) a mixed version, the lowest check node degree criterion is used to connect the first edge to a symbol node (not only to 2-degree symbol nodes as proposed here) and the FCD criterion is used for the remaining edges.

\begin{figure}[t]
\includegraphics[width=\linewidth]{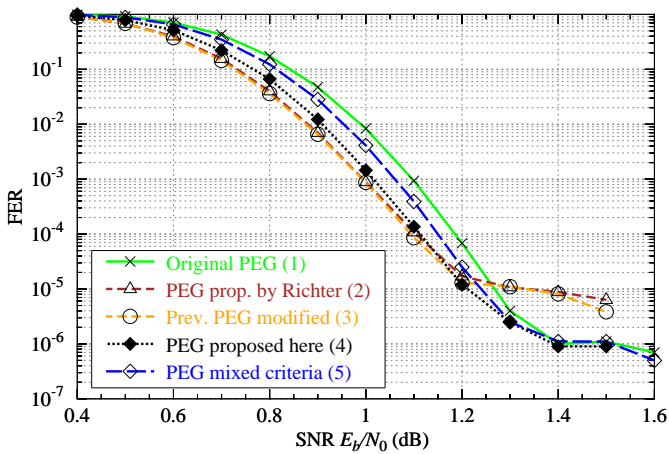}
\caption{Frame error rate (FER) as a function of the signal-to-noise ratio (SNR) for five construction methods, as explained in the text. Results have been obtained with the simulation of transmissions over the AWGN channel. Generating polynomials used in the code construction have been extracted from Table II (with a maximum symbol node degree of 50) in~\cite{Richardson01}.}
\label{fig:AWGN-SNR-FER}
\end{figure}

\begin{figure}[t]
\includegraphics[width=\linewidth]{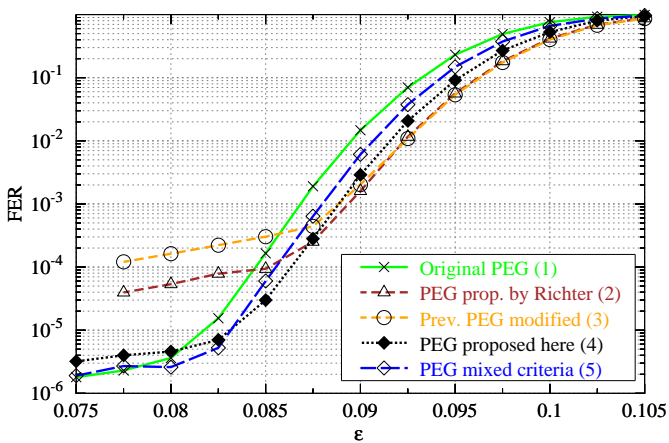}
\caption{FER as a function of the crossover probability, $\varepsilon$, over the BSC for five construction methods (see text). Codes have been constructed using the optimised generating polynomials from~\cite{Elkouss09}. Since the $\rho(x)$ distribution is more complex in this set of codes, it is possible to better appreciate the differences among the various algorithms used for constructing the codes.}
\label{fig:BSC-SNR-FER}
\end{figure}

Figs.~\ref{fig:AWGN-SNR-FER} and~\ref{fig:BSC-SNR-FER} show the performance of the codes over the additive white Gaussian noise (AWGN) channel and the BSC. The error floor is improved using the zig-zag construction for 2-degree symbol nodes. On the other hand,  within a given graph setting, when the first check node connected to a symbol node rule is used, there is no relevant improvement, as this allows for different structures to the zig-zag.

\begin{table}
\caption{$\rho$-Compliance calculated as defined in Eq.~(\ref{eq:compliance})}
\label{tab:rho-compliance}
\begin{minipage}{1.0\linewidth}
\centering
\begin{tabular}{l c c c c c}
\hline
Algorithm & (1) & (2) & (3) & (4) & (5)\\
\hline
AWGN & 1.495936 & 0.022604 & 0.016141 & 0.020870 & 0.022998 \\
BSC & 1.938722 & 0.050631 & 0.051067 & 0.057456 & 0.050050 \\
\hline
AWGN\footnote{Using relaxed edge-selection.} & -- & 0.001482 & 0.001403 & 0.001673 & 0.000806 \\
BSC & -- & 0.001623 & 0.000459 & 0.001634 & 0.000510 \\
\hline
\end{tabular}
\end{minipage}
\end{table}

\section{Conclusions}

In this paper, we proposed an improved PEG algorithm. The constructed codes comply with both an irregular check and an irregular symbol degree distribution. Simulation results show that this method behaves very well in the waterfall region while also maintaining a low error floor. Good performance in both regions is relevant as it allows to use the same code independently of the working point, e.g. this construction might find application in rate compatible solutions.

\section*{Acknowledgment}

Partially supported by project Quantum Information Technologies Madrid\footnote{http://www.quitemad.org}, P2009/ESP-1594, Co\-mu\-ni\-dad Au\-t\'o\-no\-ma de Madrid. The authors acknowledge the resources and assistance provided by the \emph{Centro de Supercomputaci\'on y Visualizaci\'on de Madrid}\footnote{http://www.cesvima.upm.es} and the Spanish Supercomputing Network.

\bibliographystyle{IEEEtran}
\bibliography{impeg}

\vfill

\end{document}